\documentclass{ws-ijmpcs}

\def\draftnote{}

\usepackage{psfrag}

\begin{document}

\newcommand{\beq}{\begin{equation}}
\newcommand{\eeq}{\end{equation}}
\newcommand{\mpi}{M_{\pi}}
\newcommand{\diff}{\text{d}}
\renewcommand{\Re}{\text{Re}\,}
\renewcommand{\Im}{\text{Im}\,}
\newcommand{\disc}{\text{disc}\,}
\newcommand{\discan}{\text{disc}_\text{an}\,}

\markboth{Martin Hoferichter, Gilberto Colangelo, Massimiliano Procura, Peter Stoffer}
{Virtual photon--photon scattering}

%
\catchline{}{}{}{}{}
%

\title{VIRTUAL PHOTON--PHOTON SCATTERING}

\author{Martin Hoferichter\footnote{E-mail address:
hoferichter@itp.unibe.ch}~, Gilberto Colangelo, Massimiliano Procura, Peter Stoffer}

\address{
Albert Einstein Center for Fundamental Physics,
Institute for Theoretical Physics,\\ 
University of Bern, Sidlerstrasse 5, CH--3012 Bern, Switzerland}

\maketitle


\begin{abstract}
Based on analyticity, unitarity, and Lorentz invariance the contribution from hadronic vacuum polarization to the anomalous magnetic moment of the muon is directly related to the cross section of $e^+e^-\to\text{hadrons}$. We review the main difficulties  
that impede such an approach for light-by-light scattering and identify the required ingredients from experiment. Amongst those, the most critical one is the scattering of two virtual photons into meson pairs. We analyze the analytic structure of the process $\gamma^*\gamma^*\to\pi\pi$ and show that the usual Muskhelishvili--Omn\`es representation can be amended in such a way as to remain valid even in the presence of anomalous thresholds.

\end{abstract}

\ccode{PACS numbers: 11.55.Fv, 13.40.Em, 13.60.Fz, 13.75.Lb}

\section{Hadronic Vacuum Polarization}

The leading contribution of strong interactions to the anomalous magnetic moment of the muon $g-2$ originates from 
hadronic intermediate states in the polarization tensor of the photon.\cite{JF} By means of gauge invariance, the polarization tensor may be expressed in terms of one single-variable scalar function $\Pi(k^2)$
\beq
 \raisebox{-0.38cm}{\includegraphics[width=3cm,clip]{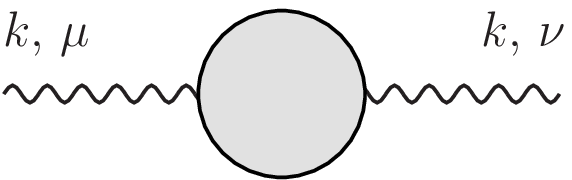}}\ =\ -i\big(k^2 g^{\mu\nu}-k^\mu k^\nu\big)\Pi\big(k^2\big).
\eeq
Due to analyticity, the renormalized self energy satisfies a subtracted dispersion relation
\beq
 \Pi_\text{ren}=\Pi\big(k^2\big)-\Pi(0)=\frac{k^2}{\pi}\int\limits_{4\mpi^2}^\infty\diff s\frac{\Im \Pi(s)}{s\big(s-k^2\big)}.
\eeq
Unitarity relates the imaginary part to the $e^+e^-$ hadronic cross section
\beq
\Im \Pi(s)=\frac{s}{4\pi\alpha}\sigma_\text{tot}\big(e^+e^-\to\text{hadrons}\big).
\eeq
In this way, general principles obeyed by the polarization tensor provide a direct link between its contribution to $g-2$ and observables.

\section{Light-by-Light Scattering}

\subsection{Structure of the Light-by-Light Tensor}

No such immediate relation to experiment is known for the light-by-light tensor $\Pi_{\mu\nu\lambda\sigma}$, describing the scattering process
\beq
\gamma^*(q_1,\mu)\gamma^*(q_2,\nu)\to\gamma^*(-q_3,\lambda)\gamma(k,\sigma).
\eeq
In contrast to vacuum polarization, there are $29$ independent Lorentz structures, cf.\ Ref.~\refcite{Bijnens95}, and $5$ independent kinematic variables ($2$ Mandelstam variables and $3$ virtualities), so that the full amplitude should be expanded in a suitable set of basis functions\footnote{As shown in Ref.~\refcite{Aldins}, gauge invariance for the on-shell photon implies that only the derivative with respect to $k_\rho$ at $k=0$ is needed for the application in $g-2$.}
\beq
 \Pi^{\mu\nu\lambda\sigma}\big(q_1,q_2,q_3\big)=\sum_{i=1}^{29}A_i^{\mu\nu\lambda\sigma}\big(q_1,q_2,q_3\big)\Pi_i\big(s,t,q_1^2,q_2^2,q_3^2\big).
\eeq
In order to write down dispersion relations for the scalar coefficients $\Pi_i$, the basis functions $A_i^{\mu\nu\lambda\sigma}$ need to be chosen in such a way that the $\Pi_i$ are free of kinematic singularities and that crossing symmetry, e.g.\ invariance under $(q_1,\mu)\leftrightarrow(q_2,\nu)$, is maintained.

\begin{figure}
\centering
 \includegraphics[width=1.8cm,clip]{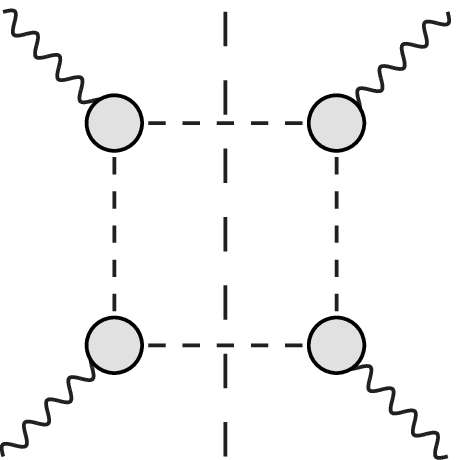}\quad
 \includegraphics[width=1.8cm,clip]{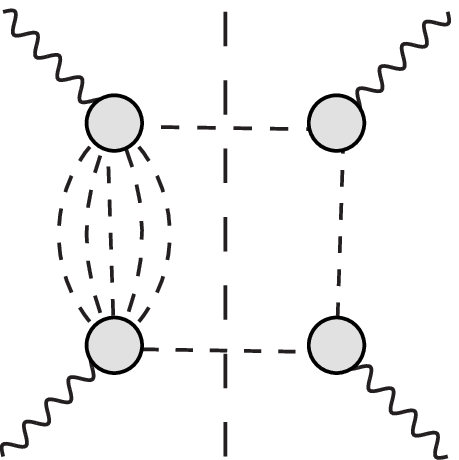}\quad
 \includegraphics[width=1.8cm,clip]{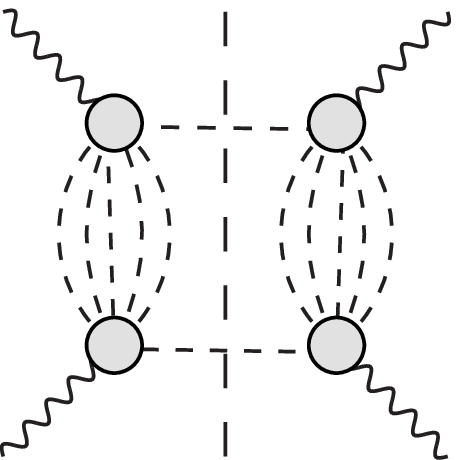}\quad
 \includegraphics[width=1.8cm,clip]{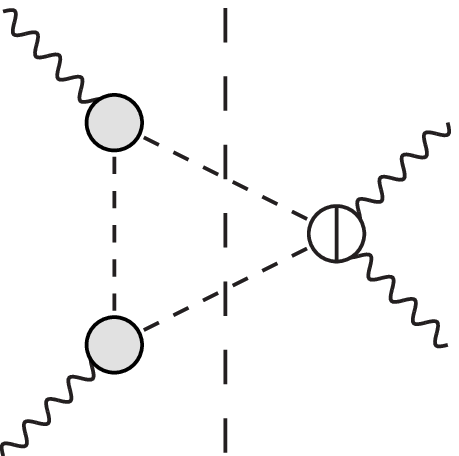}\quad
 \includegraphics[width=1.8cm,clip]{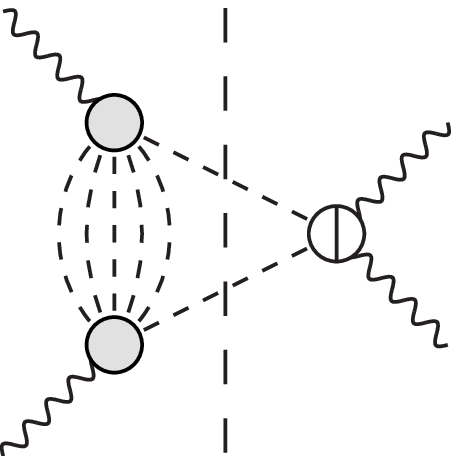}\quad
 \raisebox{0.4cm}{\includegraphics[width=1.8cm,clip]{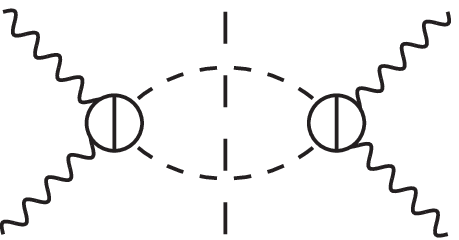}}
\caption{Classes of unitarity diagrams in light-by-light scattering. The grey blobs denote (transition) form factors, the blobs with vertical line a polynomial contribution in the crossed channel. Short-dashed lines refer to pions, wiggly lines to photons, and the long-dashed lines indicate cut propagators. Crossed diagrams are not shown.}
\label{fig:unitarity}
\end{figure}

The complicated structure of the light-by-light tensor prohibits a comprehensive analysis of all intermediate states allowed by unitarity. However, the most important states (besides the pseudoscalar meson poles) in the low/intermediate energy region are two-meson reducible. They can be classified according to the analytic structure in the crossed channel as shown in Fig.~\ref{fig:unitarity}. There are classes of box, triangle, and bulb unitarity diagrams, depending on whether the crossed-channel amplitude involves non-polynomial terms. Such non-polynomial contributions are given by the pion pole and multi-pion exchange, whereas the polynomials for instance include effects due to $\pi\pi$ rescattering. In practice, the multi-pion diagrams may be approximated by resonance exchange, i.e.\ $\rho$ and $\omega$/$\phi$ for $2$ and $3$ pions, respectively. While for $\omega$ and $\phi$ a narrow-width approximation is certainly viable, the effect of the finite width of the $\rho$ is captured through a spectral-function approach that relies on the amplitude for $\gamma^*\pi\to\pi\pi$ as input.      

\subsection{Input from Experiment}

The experimental ingredients necessary for this program follow from Fig.~\ref{fig:unitarity}. 
Diagrams with a pion pole require the pion vector form factor, those with resonance exchange the corresponding transition form factors and the $\gamma^*\pi\to\pi\pi$ amplitude. This input for the multi-pion diagrams can again be checked for consistency within a framework respecting analyticity and unitarity.\cite{HKS,NKS,SKN}
The most critical input concerns the polynomial pieces, since they involve the pole-subtracted partial waves for the process $\gamma^*\gamma^*\to\pi\pi$. Absent direct experimental information for arbitrary virtualities, e.g.\ from $e^+e^-\to \pi\pi\ell^+\ell^-$, these partial waves are again reconstructed dispersively, see Refs.~\refcite{GM10,HPS} for two on-shell photons and Ref.~\refcite{Moussallam13} for one photon with non-vanishing virtuality. Finally, the dispersion relations for the $\Pi_i$ will involve a contribution of the pion-pole diagram, with a residue determined by the (on-shell) pion transition form factor $F_{\pi\gamma^*\gamma^*}(\mpi^2,q_1^2,q_2^2)$.
In order to eliminate the model-dependence as far as possible, also input for this form factor should fulfill analyticity and unitarity requirements and be backed by data wherever available.\cite{Czerwinski,MesonNet,inprogress}

\section{Analytic Structure of $\boldsymbol{\gamma^*\gamma^*\to\pi\pi}$}

In principle, the partial waves for $\gamma^*\gamma^*\to\pi\pi$ are constrained by a similar set of dispersion relations as derived in Refs.~\refcite{GM10,HPS,Moussallam13}. Within a simplified scalar toy example, where the left-hand cut is approximated by the pion pole, one thus obtains the following Muskhelishvili--Omn\`es representation for the pole-subtracted $S$-wave
\beq
\label{Omnes}
f_0\big(s;q_1^2,q_2^2\big)=\frac{\Omega_0(s)}{\pi}\int\limits_{4\mpi^2}^\infty\diff s'\frac{N_0\big(s';q_1^2,q_2^2\big)\sin \delta_0(s')}{(s'-s)|\Omega_0(s')|},
\eeq
with the projection of the pole term
\begin{align}
\label{pole}
N_0\big(s;q_1^2,q_2^2\big)&=\frac{2L}{\sigma_s\sqrt{\lambda\big(s,q_1^2,q_2^2\big)}},\qquad
 L=\log\frac{s-q_1^2-q_2^2+\sigma_s\sqrt{\lambda(s,q_1^2,q_2^2)}}{s-q_1^2-q_2^2-\sigma_s\sqrt{\lambda(s,q_1^2,q_2^2)}},\notag\\
 \sigma_s&=\sqrt{1-\frac{4\mpi^2}{s}},\qquad \lambda(x,y,z)=x^2+y^2+z^2-2(xy+xz+yz),
\end{align}
the Omn\`es function $\Omega_0(s)$, and $\pi\pi$ $S$-wave $t_0(s)$
\beq
\Omega_0(s)=\exp\Bigg\{\frac{s}{\pi}\int\limits_{4\mpi^2}^\infty\diff s'\frac{\delta_0(s')}{s'(s'-s)}\Bigg\},\qquad 
t_0(s)=\frac{1}{\sigma_s}e^{i\delta_0(s)}\sin\delta_0(s).
\eeq
The analytic continuation of this solution in the virtualities $q_i^2$ in the case that both photons are off-shell is complicated by the occurrence of anomalous thresholds,\cite{Mandelstam} i.e.\ the singularities of the logarithm in Eq.~\eqref{pole} located at
\beq
 s_\pm=q_1^2+q_2^2-\frac{q_1^2q_2^2}{2\mpi^2}\pm\frac{1}{2\mpi^2}\sqrt{q_1^2\big(q_1^2-4\mpi^2\big)q_2^2\big(q_2^2-4\mpi^2\big)}.
\eeq
In this way, left- and right-hand cut become intertwined, which invalidates the direct derivation of Eq.~\eqref{Omnes} for large virtualities.

\begin{figure}
\centering
 \includegraphics[width=6cm,clip]{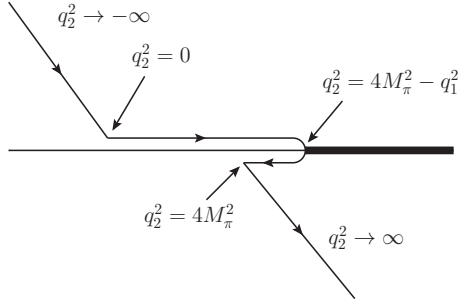}
\caption{Trajectory of the anomalous branch point $s_+$ as a function of $q_2^2$ for $0\leq q_1^2\leq 4\mpi^2$. For $q_2^2\to-\infty$, $s_+$ lies on the second sheet, then migrates onto the first sheet through the unitarity cut, and there requires a deformation of the integration contour.}
\label{fig:trajectory}
\end{figure}

\begin{figure}
\centering
\includegraphics[width=\linewidth]{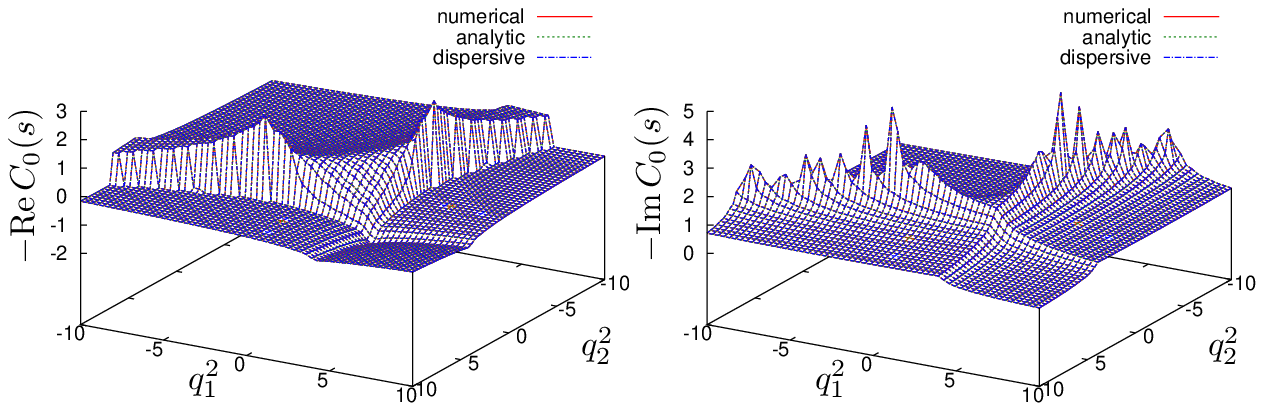}\\[3mm]
\includegraphics[width=\linewidth]{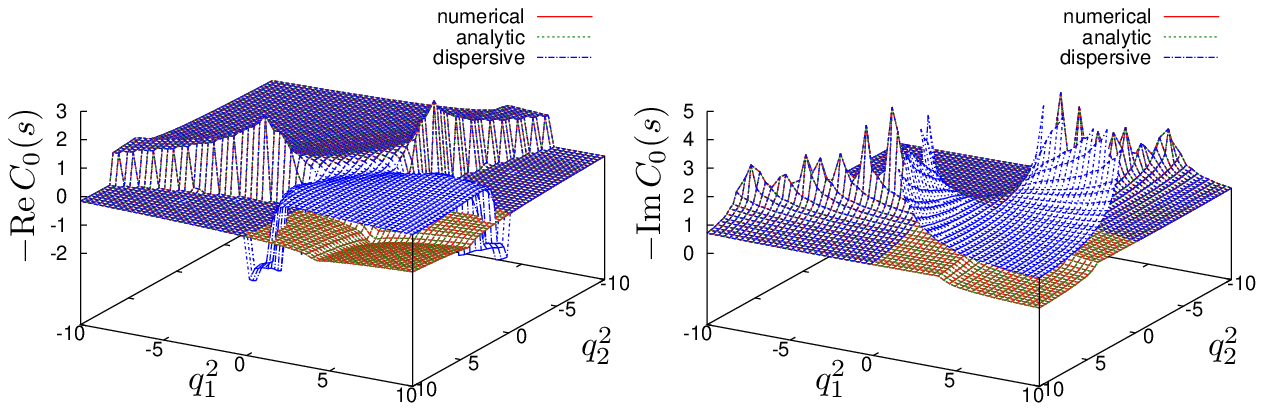}
\caption{$C_0(s)$ for $s=5$ and $\mpi=1$ calculated numerically, analytically, and dispersively. The lower panel shows the effect of switching off the anomalous contribution in the dispersive formula.}
\label{fig:C0_anom}
\end{figure}

In order to elucidate the role of these anomalous thresholds we first consider the scalar triangle loop function
\beq
C_0(s)=\frac{1}{i\pi^2}\int\frac{\diff^4k}{\big(k^2-\mpi^2\big)\big((k+q_1)^2-\mpi^2\big)\big((k-q_2)^2-\mpi^2\big)}.
\eeq
If $q_1^2+q_2^2\geq 4\mpi^2$, its dispersive representation involves an additional, anomalous piece that emerges because the anomalous branch point's moving onto the first sheet distorts the integration contour, see Fig.~\ref{fig:trajectory} and Ref.~\refcite{LMS}. The numerical results in Fig.~\ref{fig:C0_anom} show that the dispersive reconstruction of $C_0(s)$ indeed works for arbitrary virtualities as long as the anomalous contribution is taken into account (upper panel), but that substantial deviations occur in the region of large virtualities if the anomalous piece is ignored (lower panel). 

In fact, this procedure to perform the analytic continuation in the $q_i^2$ for $C_0(s)$ transfers immediately to $f_0(s;q_1^2,q_2^2)$, the crucial observation being that the integrand of Eq.~\eqref{Omnes} coincides with the discontinuity of $C_0(s)$,
\beq
 \frac{N_0\big(s;q_1^2,q_2^2\big)\sin \delta_0(s)}{|\Omega_0(s)|}=
 -\frac{\disc C_0(s)}{\pi i\sigma_s}\frac{\sin \delta_0(s)}{|\Omega_0(s)|}
 =-\frac{\disc C_0(s)}{\pi i}\frac{t_0(s)}{\Omega_0(s)},
 \eeq
 up to a factor $t_0(s)/\Omega(s)$, which is independent of $q_i^2$ and well-defined in the whole complex $s$-plane. Therefore, the full result for $f_0(s;q_1^2,q_2^2)$ becomes merely amended by an additional term that takes care of the anomalous thresholds  
\begin{align}
   f_0\big(s;q_1^2,q_2^2\big)\Big|_\text{anom}&=
   \theta\big(q_1^2+q_2^2-4\mpi^2\big)\frac{\Omega_0(s)}{2\pi i}\int\limits_0^1\diff x\frac{\partial s_x}{\partial x}\frac{\discan f_0\big(s_x;q_1^2,q_2^2\big)}{s_x-s},\notag\\
   \discan f_0\big(s;q_1^2,q_2^2\big)&=-\frac{8\pi}{\sqrt{\lambda\big(s,q_1^2,q_2^2\big)}}\frac{t_0(s)}{\Omega_0(s)},\qquad
   s_x=4\mpi^2\,x+(1-x)s_+.\notag
\end{align}

\section*{Acknowledgments}

We would like to thank Bastian Kubis, Bachir Moussallam, and Sebastian Schneider for numerous useful discussions.
Financial support by the Swiss National Science Foundation is gratefully acknowledged.
The AEC is supported by the
  ``Innovations- und Kooperationsprojekt C-13'' of the ``Schweizerische
  Universit\"atskonferenz SUK/CRUS.''

\end{document}